\pdfoutput=1

\documentclass[aps,preprintnumbers,amsmath,amssymb,twocolumn, tightenlines,superscriptaddress]{revtex4}

\usepackage{graphicx}
\usepackage{epsfig}
\usepackage{float}
\usepackage{color}
\usepackage{ulem}

\newcommand{\be}{\begin{eqnarray}}
\newcommand{\ee}{\end{eqnarray}}

 \newcommand{\gsim}{\mathrel{\hbox{\rlap{\lower.55ex \hbox {$\sim$}}
                   \kern-.3em \raise.4ex \hbox{$>$}}}}
\newcommand{\lsim}{\mathrel{\hbox{\rlap{\lower.55ex \hbox {$\sim$}}
                   \kern-.3em \raise.4ex \hbox{$<$}}}}

\newcommand{\ba}{\begin{eqnarray}}
\newcommand{\ea}{\end{eqnarray}}

\begin{document}


\title{ Dynamical induced quark spin polarization by magnetic field at the early stage of heavy-ion collisions}
\author{Anping Huang} \email{huanganping@ucas.ac.cn}
\address{School of Material Science and Physics, China University of Mining and Technology, Xuzhou, China}
\address{School of Nuclear Science and Technology, University of Chinese Academy of Sciences, Beijing 100049, China}
\author{Zilin Yuan} 
\address{School of Nuclear Science and Technology, University of Chinese Academy of Sciences, Beijing 100049, China}
\author{Mei Huang} \email{huangmei@ucas.ac.cn}
\address{School of Nuclear Science and Technology, University of Chinese Academy of Sciences, Beijing 100049, China}
\date{\today}

\begin{abstract}
We present a comprehensive analysis of the dynamic process of quark spin polarization induced by magnetic fields at the pre-thermal stage in heavy-ion collisions by using the recently developed theoretical tool of chiral kinetic theory. Our findings demonstrate that the spin polarization of quarks is highly sensitive to the interactions between quarks. These interactions can delay the decay of early spin polarization vector while accelerating the decay of later spin polarization vector. Specifically, our simulations show the detailed process of how magnetic fields polarize quarks within the fireball and reveal that quark interactions lead to an acceleration effect on the average spin. Notably, the fireball of quark-gluon plasma (QGP) in its early stages exhibits an incomplete electromagnetic response effect, which differs from the response predicted by Lenz's law. This discrepancy arises from quantum corrections involving the interactions between quark spin and electromagnetic fields.

\end{abstract}

\maketitle

\section{Introduction}

The spin degree of freedom is a fundamental characteristic of modern physics that has numerous applications in various fields. In particular, in heavy-ion collisions, it has been attracted considerable attention to studying many physical phenomena related to spin, especially the quark matter under external magnetic field and vortical field, such as the chiral magnetic effect ~\cite{Kharzeev:2004ey,Kharzeev:2007tn,Kharzeev:2007jp,Fukushima:2008xe}, chiral magnetic wave~\cite{Kharzeev:2010gd,Burnier:2011bf}, Chiral vortical effects~\cite{Son:2009tf,Kharzeev:2010gr,Sadofyev:2010is,Landsteiner:2011iq,Jiang:2015cva}, and spin polarization effect~\cite{STAR:2017ckg, STAR:2018gyt,STAR:2019erd,Niida:2018hfw,ALICE:2019onw}, etc. 

The spin polarization of $\Lambda $ and $\bar{\Lambda}$ hyperons have been observed in 7.7- 200 A GeV Au + Au collisions through the angular distribution of their weak-decay products~\cite{STAR:2017ckg, STAR:2018gyt, STAR:2019erd, Niida:2018hfw}. The global spin polarization effect has been successfully described by the thermal vorticity effects~\cite{Becattini:2013fla, Fang:2016vpj, Pang:2016igs}. Recently, the “spin
sign puzzle”~\cite{Liu:2019krs, Wu:2019eyi, Florkowski:2019voj} of the local $\Lambda$ polarization also has been explained by the shear-induced polarization (SIP) effect~\cite{Fu:2021pok,Fu:2020oxj}. Despite extensive research, the splitting of polarization between $\Lambda $ and $\bar{\Lambda}$ hyperons remains an open question~\cite{Muller:2018ibh,Guo:2019mgh,Guo:2019joy,Han:2019fce,Becattini:2020ngo, Xu:2022hql}. Recently, the strong magnetic fields present in heavy-ion collisions have been proposed to understand the polarization splitting puzzle in $\Lambda$ and $\bar{\Lambda}$ hyperons \cite{Xu:2022hql} .

There has been significant interest in recent years in studying the spin polarization of hyperons in heavy-ion collisions. However, describing the detailed dynamic spin polarization of quarks in quark-gluon plasma fireballs remains a significant challenge. The spin polarization of $\Lambda$ hyperons in heavy-ion collisions is believed to be encoded in the behavior of quarks in quark-gluon plasma (QGP), particularly the strange quark~\cite{Fu:2021pok,Fu:2020oxj}. Therefore, understanding the behavior of the strange quark in QGP is the key to unraveling the spin-splitting puzzle in $\Lambda$ hyperons. At present, the quantitative modeling of spin polarization of $\Lambda$ hyperons is quantitatively modeled only at the freeze-out surface of hydrodynamics using the spin Cooper-Frye formula. However, an  unresolved crucial issue is the detailed evolution of the spin polarization of quarks in QGP. 

In this study, we take an important step towards addressing this challenge by using the recently developed theoretical tool of chiral kinetic theory~\cite{Stephanov:2012ki, Son:2012wh, Son:2012zy, Chen:2012ca, Kharzeev:2016sut, Chen:2012ca, Hidaka:2016yjf, Mueller:2017arw, Gorbar:2017cwv, Huang:2018wdl} to investigate the dynamic evolution of the spin splitting between quarks and anti-quarks induced by magnetic field at the pre-thermal stage of heavy-ion collisions. At this stage, the quark-gluon plasma is far from thermal equilibrium and the magnetic field is at its strongest. In the chiral limit, the properties of the d and s quarks are identical, including their charge. As a result, for our simulation, we will only need to consider the u and d quarks. 

The paper is organized as following: after introduction, we will introduce the theoretical framework in Sec.\ref{sec-frame}, and we present numerical method and results in Sec.\ref{numrical} and Sec.\ref{resuts}, and we give a summary in Sec. \ref{summary}.

\section{The theoretical framework }
\label{sec-frame}
The theoretical framework to describe the transport phenomenon of chiral quarks under electromagnetic fields in an out-of-equilibrium system is composed of the chiral kinetic theory and Maxwell equations~\cite{Stephanov:2012ki, Son:2012wh, Son:2012zy, Chen:2012ca, Kharzeev:2016sut, Hidaka:2016yjf, Mueller:2017arw, Gorbar:2017cwv, Huang:2018wdl}. The chiral kinetic equation is to take the following form~\cite{Stephanov:2012ki, Chen:2015gta, Huang:2017tsq, Huang:2018wdl}:
\begin{align}\label{eq:ckt}
\begin{split}
&\Bigg\{ \partial_{t}+\dot{\mathbf{x}}\cdot\triangledown_{\mathbf{x}}+\dot{\mathbf{p}}\cdot\triangledown_{\mathbf{p}}\Bigg\}f^{i}_{\chi}(x,\mathbf{p})=C[f^{i}_{\chi}].
\end{split}
\end{align}
with
\begin{align}\label{eq:eom}
\begin{split}
&\dot{\mathbf{x}}=\frac{1}{\sqrt{G}}\left(\widetilde{\boldsymbol{v}}+\hbar q_{i}(\widetilde{\boldsymbol{v}}\cdot\mathbf{b}_{\chi})\mathbf{B}+\hbar q_{i}\widetilde{\mathbf{E}}\times\mathbf{b}_{\chi}\right),\\
&\dot{\mathbf{p}}=\frac{q_{i}}{\sqrt{G}}\left(\widetilde{\mathbf{E}}+\widetilde{\boldsymbol{v}}\times\mathbf{B}+\hbar q_{i}(\widetilde{\mathbf{E}}\cdot\mathbf{B})\mathbf{b}_{\chi} \right).
\end{split}
\end{align}
Where the corresponding Jacobian, energy, group velocity are
\begin{align*}
&\sqrt{G}=\left(1+\hbar q_{i}\mathbf{b}_{\chi}\cdot\mathbf{B} \right),\\
&\widetilde{\mathbf{E}}=\mathbf{E}-\frac{1}{ q_{i}}\triangledown_{\mathbf{x}}E_{\mathbf{p}},~~~E_{\mathbf{p}}=|\mathbf{p}|\left( 1-\hbar q_{i}  \mathbf{B}\cdot \bf{b}_{\chi}\right),\\
&\widetilde{\boldsymbol{v}}=\frac{\partial E_{\mathbf{p}}}{\partial\mathbf{p}}=\widehat{\mathbf{p}}\left(1+2\hbar q_{i}\mathbf{B}\cdot\mathbf{b}_{\chi} \right)-\hbar q_{i} b_{\chi}\mathbf{B}.
\end{align*}
Where $q_i$ is the electric charge of the chiral fermion, $\chi$ the chirality, and $\mathbf{b}_{\chi}=\chi\frac{\mathbf{p}}{2|\mathbf{p}|^{3}}$ the Berry curvature.  
While $f^{i}_{\chi}$ is the distribution function in the phase space of position $\bf x$ and momentum $\bf p$ for each specie (labelled by $i$) of chiral fermions,  and  $C[f^{i}_{\chi}]$ is the collision term. The collision term in chiral kinetic theory is more complex than that in classical kinetic theory, and it is corrected by quantum corrections. For a more detailed explanation, please refer to the works\cite{ Gorbar:2016qfh, Hidaka:2016yjf, Yang:2020hri, Yamamoto:2023okm}.   

The companied Maxwell equations can be written as the following,
\begin{align}\label{eq:me}
\begin{split}
&\triangledown\cdot\mathbf{E}=\rho_{e},\\
&\triangledown\cdot\mathbf{B}=0,\\
&\partial_{t}\mathbf{E}=\triangledown\times\mathbf{B}-\mathbf{j}_{e},\\
&\partial_{t}\mathbf{B}=-\triangledown\times\mathbf{E}.
\end{split}
\end{align} 
Where $\rho_{e}$ and $\mathbf{j}_{e}$ are the charge and charge current density, respectively. In heavy-ion collisions, they can be divided into two parts as  $\rho_{e}=\rho^{\text{ext}}_{e}+\rho^{\text{int}}_{e}$ and $\mathbf{j}_{e}=\mathbf{j}^{ext}_{e}+\mathbf{j}^{int}_{e}$ \cite{mclerran2014comments, Huang:2022qdn}. Herein, $\rho^{\text{ext}}_{e}$ and $\mathbf{j}^{ext}_{e}$ are the contribution coming from an external source like the fast-moving charged particles, which are the protons from colliding nuclei. While $\rho^{\text{int}}_{e}$ and $\mathbf{j}^{int}_{e}$ are contributions from the medium, which can be derived by the chiral distribution function as following,
\begin{align}\label{eq:ec}
\begin{split}
&\rho^{int}_{e}=\sum_{i}q_{i}\sum_{\chi=\pm} \int \frac{d^3p}{(2\pi)^3} \sqrt{G} f^{i}_{\chi}(x,\mathbf{p}),\\
&\mathbf{j}^{int}_{e}=\sum_{i}q_{i}\sum_{\chi=\pm} \int \frac{d^3p}{(2\pi)^3} \sqrt{G} \dot{\mathbf{x}} f^{i}_{\chi}(x,\mathbf{p}).
\end{split}
\end{align}
Herein the species $i=(u, \bar{u}, d, \bar{d})$, with the electric charge $q_u=-q_{\bar{u}}=\frac{2e}{3}$ and $q_d=-q_{\bar{d}}=-\frac{e}{3}$. It is explicit that the internal electric density and current density are the bridge between the chiral transport equations and Maxwell's equations. The manifestation of the quantum chiral anomaly effect within the framework of Maxwell's equations is evident through these two quantities. For instance, this effect can be observed by performing momentum integration on the electric current density, utilizing the relation provided in the first line of Eq.(\ref{eq:eom}), like as the following,
\begin{align}
	\begin{split}
		&\mathbf{j}^{int}_{e}\simeq \mathbf{j}_{normal}+\mathbf{j}_{ahc}+\mathbf{j}_{cme}.\\
		&\mathbf{j}_{normal}=\sum_{i}q_{i}\sum_{\chi=\pm} \int \frac{d^3p}{(2\pi)^3} f^{i}_{\chi} \hat{\mathbf{p}},\\
		&\mathbf{j}_{ahc}= \hbar\mathbf{E}\times\sum_{i}q_{i}\sum_{\chi=\pm} \int \frac{d^3p}{(2\pi)^3} f^{i}_{\chi} \mathbf{b}_{\chi},\\
		&\mathbf{j}_{cme}= \hbar\mathbf{B}\sum_{i}q_{i}\sum_{\chi=\pm} \int \frac{d^3p}{(2\pi)^3} f^{i}_{\chi} \hat{\mathbf{p}}\cdot\mathbf{b}_{\chi}.
	\end{split}
\end{align}
The first term represents the normal current, the second term corresponds to the anomalous Hall current, and the final term accounts for the chiral magnetic effect (CME) current.

In this work, we will focus on the dynamical spin polarization of the quarks $\mathbf{P}_{i}$ in QGP, which can be quantified by the axial current scaled by the number of the quark \cite{Han:2017hdi, Liu:2019krs}, 
\begin{align}\label{eq:sp}
\begin{split}
&\mathbf{P}_{i}(t)=\frac{\int d^3x\, \mathbf{j}_{5,i}(t,\mathbf{x})}{\int d^3x\, n_{i}(t,\mathbf{x})}\\
&~~~~~~~~=\frac{\int d^3x \left(\mathbf{j}_{i,R}(t,\mathbf{x})- \mathbf{j}_{i,L}(t,\mathbf{x})\right)}{\int d^3x \left( n_{i,R}(t,\mathbf{x})+ n_{i,L}(t,\mathbf{x})\right)}\\
&~~~~~~~~=\frac{\mathbf{J}_{i,R}(t)- \mathbf{J}_{i,L}(t)}{N_{i,R}+N_{i,L}}.
\end{split}
\end{align}
This definition differs from those presented in \cite{Becattini:2013fla, Becattini:2021suc, Becattini:2020sww, Fu:2021pok}, which pertain to massive particles. In fact, these two definitions are essentially equivalent, as they both derive from the Pauli–Lubanski vector; however, they differ in how the mean spin vector is defined. For massive particles, this vector is obtained by dividing the Pauli-Lubanski vector by the mass, whereas for massless particles, it is divided by the energy. Consequently, the spin polarization can be calculated through the following procedure: the mean spin vector $\bar{S}^{\mu}(x,p)=\frac{1}{2p\cdot u}\epsilon^{\mu\nu\rho\sigma}J_{\nu\rho}p_{\sigma}/\frac{dN}{\frac{d^3xd^3p}{(2\pi)^3}}=-\frac{1}{2}A^{\mu}(x,p)/\frac{dN}{\frac{d^3xd^3p}{(2\pi)^3}}=\frac{1}{2}J^{\mu}_5(x,p)/\frac{dN}{\frac{d^3xd^3p}{(2\pi)^3}}$, the total mean spin vector is $\bar{S}^{\mu}=\int \frac{d^3xd^3p}{(2\pi)^3}\bar{S}^{\mu}(x,p)/N=\frac{1}{2}J^{\mu}_5/N$, and the total global spin polarization is defined as $P^{\mu}=\bar{S}^{\mu}/s=J^{\mu}_5/N$. Here, $s=1/2$, and N is the number of paritcle. The axial vector $A^{\mu}$,is related to the axial current by $J^{\mu}_5=-A^{\mu}$. The relevant derivations can be referenced in the works\cite{Liu:2021nyg, Becattini:2020sww, Fang:2024vds}. The number and chiral current of quarks are expressed by the following, 
\begin{align}\label{eq:cc}
\begin{split}
N_{i,\chi}(t)&=\int \frac{d^3pd^3x}{(2\pi)^3} \sqrt{G} f^{i}_{\chi}(x,\mathbf{p}),\\
\mathbf{J}_{i,\chi}(t)&=\int d^{3}x \mathbf{j}_{i,\chi}(t,\mathbf{x})\\
&=\int \frac{d^3pd^3x}{(2\pi)^3} \sqrt{G} \dot{\mathbf{x}} f^{i}_{\chi}(x,\mathbf{p}).
\end{split}
\end{align}
Where $\chi=\pm$ or $\chi=R/L$ denotes the right/left-hand quark. While another vital quantity to describe the dynamical evolution of spin polarization of quarks is the average spin vector, defined as the following,
\begin{align}\label{eq:as}
\mathbf{S}_{i}(t)&=\int d^{3}x \mathbf{S}_{i}(t,\mathbf{x})\nonumber\\
&=\sum_{\chi=\pm}\frac{1}{N_{i}}\int \frac{d^3pd^3x}{(2\pi)^3} \frac{\chi}{2} \hat{\mathbf{p}} f^{i}_{\chi}(x,\mathbf{p}).
\end{align} 
Where $N_{i}=N_{i,R}+N_{i,L}$ is the total number of species $i$ particle, and herein $\mathbf{S}_{i}(t,\mathbf{x})$ the average spin density vector. In fact, by comparing this equation with Eq.(\ref{eq:sp}), it is clear that $2\mathbf{S}_{i}(t)$ is the zeroth order part of the spin polarization defined in Eq.(\ref{eq:sp}), which can be seen by combining Eq.(\ref{eq:eom}) and Eq.(\ref{eq:cc}) into the Eq.(\ref{eq:sp}).

\section{The Strategy for numerical calculation}
\label{numrical}
To consistently solve the chiral transport equation Eq.(\ref{eq:ckt}) and Maxwell's equations Eq.(\ref{eq:me}), we choose split method to solve the chiral transport equation\cite{aristov2001direct}, and select the Yee-grid algorithm to solve Maxwell's equation\cite{yee1966numerical, mclerran2014comments, Huang:2022qdn}. 

We show the detailed algorithm for the simulation. Firstly, the chiral transport equation Eq.(\ref{eq:ckt}) is divided into the free-streaming transport equation and collision equation in a time interval $[t,t+dt]$ like the following
\begin{align}\label{eq:ckt-num}
\begin{split}
&\Bigg\{ \partial_{t}+\dot{\mathbf{x}}\cdot\triangledown_{\mathbf{x}}+\dot{\mathbf{p}}\cdot\triangledown_{\mathbf{p}}\Bigg\}f^{i}_{1\chi}(x,\mathbf{p})=0.\\
&f^{i}_{1\chi}(t)=f^{i}_{\chi}(t).\\
&\partial_{t}f^{i}_{2\chi}(x,\mathbf{p})=C[f^{i}_{2\chi}],\\
&f^{i}_{2\chi}(t)=f^{i}_{1\chi}(t+dt).
\end{split}
\end{align}
Herein, the distribution function at time $t$ is the initial condition of the free transport equation, the solution of the free-streaming transport equation at next time $t+dt$ is the initial condition of the collision equation. The solution of the collision equation at $t+dt$ is the physical distribution function at $t+dt$, i.e. $f^{i}_{\chi}(t+dt)=f^{i}_{2\chi}(t+dt)$. This method is now widely used to simulate the transport equation. For the free transport equation, one can choose different numerical methods to further calculate, such as the finite difference method, finite volume method, and test particle method\cite{bird1994molecular,grigoryev2012numerical,aristov2001direct}. In this work, we take the finite difference method. 

For Maxwell's equations we will firstly take the split method used by McLerran and Skokov\cite{McLerran:2013hla} to separate Maxwell equations into the external part and internal part before further numerical calculation. In which, the electric and magnetic fields are separated into two pieces, i.e 
\begin{align}\label{eq:sumEB}
&\mathbf{E}=\mathbf{E}_{ext}+\mathbf{E}_{int},~~~\mathbf{B}=\mathbf{B}_{ext}+\mathbf{B}_{int}.
\end{align} 
Herein, "ext" denotes the external part which is originated by the source contribution, such as the fast moving charge particles in heavy-ion collisions, while the "int" regards to the induced internal electromagnetic fields in the medium of quark-gluon plasma (QGP). The Maxwell equations in Eq.(\ref{eq:me}) now can been split into two parts. For the "external" part,
\begin{align}\label{eq:meex}
\begin{split}
&\triangledown\cdot\mathbf{E}_{ext}=\rho^{\text{ext}}_{e},\\
&\partial_{t}\mathbf{E}_{ext}=\triangledown\times\mathbf{B}_{ext}-\mathbf{J}^{\text{ext}}_{e},\\
&\triangledown\cdot\mathbf{B}_{ext}=0,\\
&\partial_{t}\mathbf{B}_{ext}=-\triangledown\times\mathbf{E}_{ext}.
\end{split}
\end{align}
Herein, $\rho^{\text{ext}}_{e}$ and $\mathbf{J}^{\text{ext}}_{e}$ are the external source contributions, which are from the fast-moving charged particles, mainly the protons of colliding nuclei. They can be written as
\begin{align}
&\rho^{\text{ext}}_{e}=e\sum_{i}\delta(\mathbf{x}_{\perp}-\mathbf{x}^{'}_{\perp,i})\delta(z-z^{'}_{i}-\beta t),\nonumber\\
&\mathbf{J}^{\text{ext}}_{e}=e\sum_{i}\beta\hat{z}\,\delta(\mathbf{x}_{\perp}-\mathbf{x}^{'}_{\perp,i})\delta(z-z^{'}_{i}-\beta t).
\end{align} 
Since the solutions are the electric and magnetic fields induced by the fast moving charged particles, the solutions of the set of equations (\ref{eq:meex}) can be obtained by boosting the electric field of the protons both in projectile and target as Ref\cite{mclerran2014comments,Zakharov:2014dia, Huang:2022qdn}. The distribution of
protons both in projectile and target can be modeled by the Woods-Saxon distribution with the standard parameters\cite{alver2008phobos}. While the "internal" part of Maxwell's equation (\ref{eq:me}) is,
\begin{align}\label{eq:meint}
\begin{split}
&\triangledown\cdot\mathbf{E}_{int}=\rho^{int}_{e},\\
&\partial_{t}\mathbf{E}_{int}=\triangledown\times\mathbf{B}_{int}-\mathbf{J}^{int}_{e},\\
&\triangledown\cdot\mathbf{B}_{int}=0,\\
&\partial_{t}\mathbf{B}_{int}=-\triangledown\times\mathbf{E}_{int}.
\end{split}
\end{align}
Where the internal density $\rho^{int}_{e}$ and current density $\mathbf{J}^{int}_{e}$ are of the electric charge density and electric current density of the QGP, respectively, which are given by the following equations. 
\begin{align}
\begin{split}
&\rho^{int}_{e}=\sum_{i}q_{i}\sum_{\chi=\pm} \int \frac{d^3p}{(2\pi)^3} \sqrt{G} f^{i}_{\chi}(x,\mathbf{p}),\\
&\mathbf{j}^{int}_{e}=\sum_{i}q_{i}\sum_{\chi=\pm} \int \frac{d^3p}{(2\pi)^3} \sqrt{G} \dot{\mathbf{x}} f^{i}_{\chi}(x,\mathbf{p}).
\end{split}
\end{align}
We will select the Yee-grid algorithm\cite{Yee:1966} to solve this set of equations to get the internally induced electric and magnetic fields. Subsequently, we will incorporate the total electric and magnetic fields, as summarized in Eq.(\ref{eq:sumEB}) into the chiral transport equation Eq.(\ref{eq:ckt-num}), to obtain the updated distribution function at the next time step.   


With the above numerical method, we now apply it to study the spin polarization of quarks that are generated during the early moments in heavy ion collisions when the created QGP is still out-of-equilibrium while the magnetic field is the strongest. The created QGP at early time is characterized by the saturation scale $Q_s$, it is on the order of $1\sim 3\rm GeV$ for RHIC and the LHC~\cite{Kowalski:2007rw}. We will take $Q_s\simeq2\rm GeV$ in this work. According to the works \cite{blaizot2012bose,blaizot2013gluon,blaizot2014quark}, the physical picture of the early stage in heavy-ion collisions is that the system is initially gluon-dominated on the time interval $0\sim 1/Q_{s}$, the quarks are generated quickly on a time scale $t\sim 1/Q_s$ and then evolve toward thermal equilibrium. In this work, we take the formation time $t_{in}=0.1{\rm fm/c}$ as the starting time of the evolution of quarks and  $t_f=0.6 {\rm fm/c}$ the end of time which is on the order of onset time for hydrodynamic evolution in heavy ion collisions.

We take the same quark initial distributions $f_0$ at $t=t_{in}$ with the previous work\cite{Huang:2017tsq}:
\begin{eqnarray}
f_0 = n_{ 0 } \, {\cal F}\left(|\vec{\bf p}|\right) \,  \exp\left[ -\frac{x^2}{R_x^2}   -\frac{y^2}{R_y^2}  -\frac{z^2}{R_z^2} \right],
\end{eqnarray}
Here, there is no chirality imbalance, the number of right-handed and left-handed quarks is set to be the same. The spatial distribution is set to be Gaussian, with three width parameters, $R_z=1/(2Q_s)$, $R_x\to (R_A-b/2)$ and $R_y \to \sqrt{R_A^2-(b/2)^2}$. Herein, the transverse widths ($R_x, R_y$) are determined by the nuclear geometry with nuclear radius $R_A$ and the impact parameter $b$.  We choose the Fermi-Dirac-like form to simulate the initial momentum distribution ${\cal F}$, i.e
\begin{align}
&{\cal F}_{FD} = \frac{1}{e^{(p-Q_s)/\Delta}+1},~~~\text{ with $\Delta=0.2\rm GeV$}
\end{align}
The overall parameter $n_0$ is fixed by normalizing the quark number density at the fireball center via $\int_{\vec{\bf p}} f_0(\vec{\bf p},x=y=z=0) \to \xi Q_s^3$ \cite{blaizot2014quark} and the parameter $\xi$ will be varied in a reasonable range to be consistent with that of typical initial condition used for hydrodynamic simulations. 


In this work, the collision term is chosen as the familiar relaxation time approximation (RTA) method for simplicity, 
\begin{align}
&C[f^{i}]\approx-\frac{f^{i}-f^{i}_{eq}}{\tau_R}.
\end{align}
Where the local equilibrium distribution function $f_{eq}=1/(e^{(\epsilon-\mu)/T}+1)$ should be determined self-consistently at any space-time point during the evolution, herein the effective local equilibrium parameters $T(t,\mathbf{x})$ and $\mu(t,\mathbf{x})$ can be fixed by the matching conditions for the energy density and number density, i.e $e_{eq}=e$ and $n_{eq}=n$. 

For the rest of the paper we will focus on the case of impact parameter $b=8.87\rm\, fm$ corresponding roughly to $20-50\%$ centrality class. One should notice that the external electric and magnetic fields are controlled by the impact parameter by fixing the center of the projectile and target on the coordinates $(-b/2,0,0)$ and $(b/2,0,0)$ respectively at the time $t=0\rm\, fm$. We use a spatial volume of $31\times 31\times 81\rm\,fm^3$ with grid size $(dx,dy,dz)=(0.8,0.8,0.035)$ fm and a finite time step $\Delta t= 0.002\rm\,fm/c$ when we solve the kinetic equations Eq.(\ref{eq:ckt-num}) and Maxwell equations Eq.(\ref{eq:meint}), while the momentum volume is set to be $26\times 26\times 26\rm\,fm^{-3}$ with grid size $(dp_x,dp_y,dp_z)=(2, 2, 2)\rm\,fm^{-1}$. In our numerical simulations, the total number of quarks and the total energy are nearly conserved between the start time $0.1$fm and the end time $0.6$fm with a very small change. One key point to note is that in our calculations, we omit all quantum corrections introduced by Berry terms in the equations of motion (Eq.\ref{eq:eom}) to avoid the divergence of the Berry curvature in the small momentum regime. This omission occurs when the condition $|q_{i}\mathbf{b}_{\chi}\cdot\mathbf{B}|>0.5$ is satisfied. Under these circumstances, which correspond to quarks with low momentum, these quarks exhibit behavior akin to that of classical particles.

\section{Results and Discussions}
\label{resuts}
\begin{figure}[!hbt]
	\begin{center}
		\includegraphics[scale=0.5]{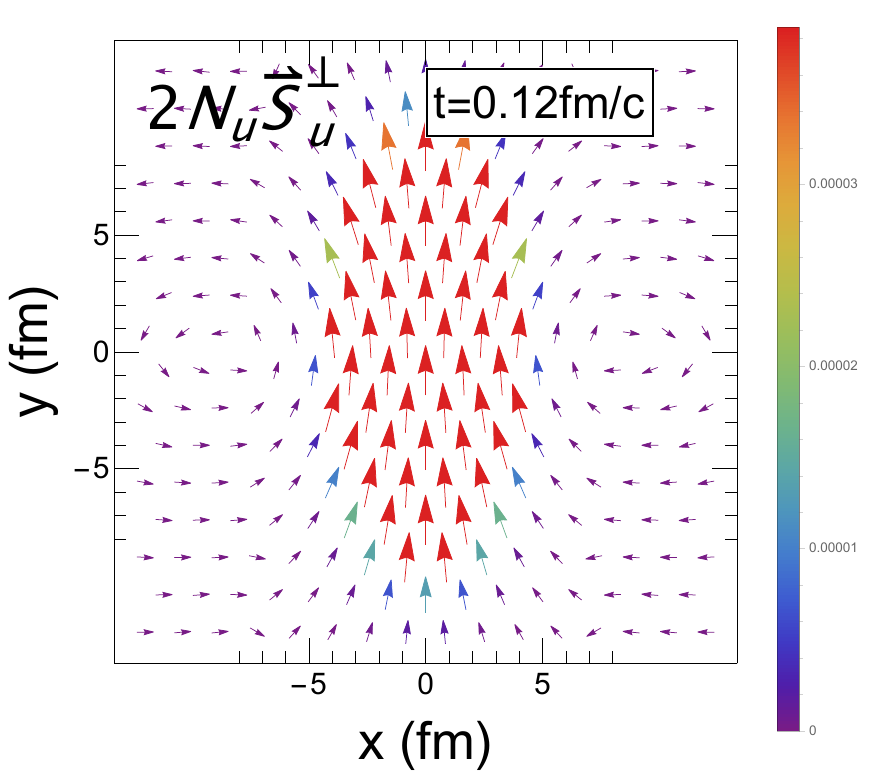}
		\caption{Transverse spin density of u quark $2N_{u}\vec{S}^{\perp}_{u}$ on the $x-y$ plane with $z=0$ at time $t = 0.12 \rm fm/c$ choosing the RTA $\tau_{R}=0.2\rm fm/c$.}
		\vspace{-0.5cm}
		\label{fig_jy}
	\end{center}
\end{figure}

Let us firstly demonstrate the out-of-equilibrium spin polarization of the quarks induced by the magnetic field by examining the transverse component $\vec{S}_{f}^\perp$ of the spin average density on the $x-y$ plane at $z=0$, where $\vec{S}_{f}^\perp=\left(S_{f}^{x},S_{f}^{y}\right)$ with $S_{f}^{x/y}=S_{f_R}^{x/y}-S_{f_L}^{x/y}$, $R/L$ denotes the right/left-handed quark and $f$ the flavor of quark. In Fig.~\ref{fig_jy} we show the $2N_{u}\vec{S}_{u}^\perp$ of u quark at time $t=0.12\rm fm/c$ with the arrow indicating the direction of the spin: it is explicitly that the spin of quark is aligned with the magnetic field and the magnitude is bigger in the area with a larger local quark density and magnetic field. While for the $\bar{u}$ quark, the results are the same except the direction is reversed. 


\begin{figure}[!hbt]
\begin{center}
\includegraphics[scale=0.32]{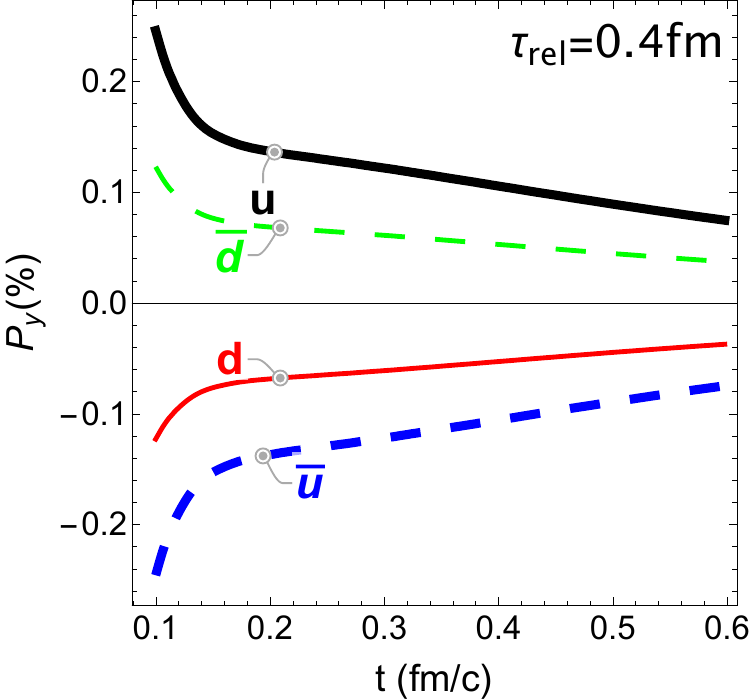}
\includegraphics[scale=0.32]{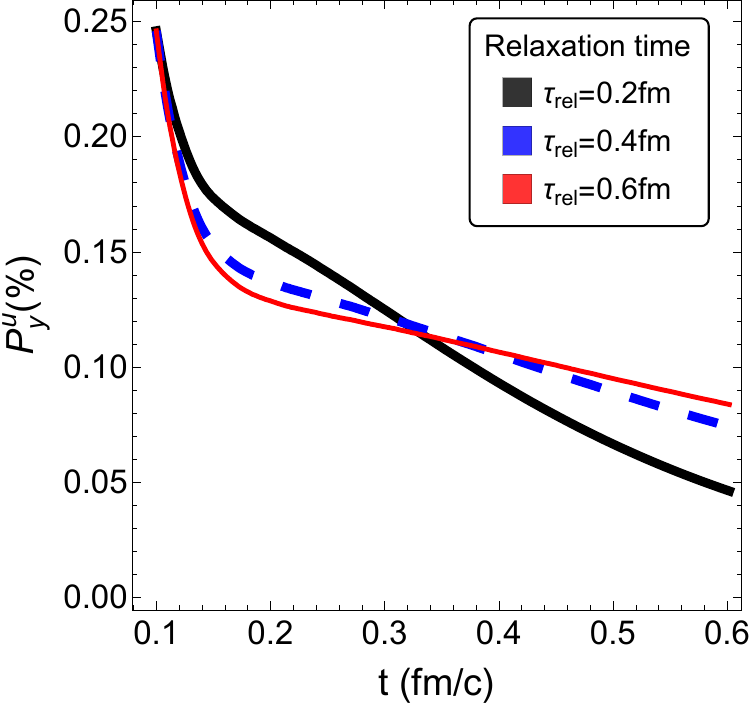}
\caption{(left)The spin polarization of different quarks as a function of time, computed with the relaxation time $\tau_{R}=0.4\mathrm{fm}$. (right) Comparing the spin polarization of u quark for different relaxation time, $\tau_{R}=0.2, 0.4, 0.6\mathrm{fm}$.}
\vspace{-0.5cm}
\label{fig_SP}
\end{center}
\end{figure}

Let's next quantify the out-of-equilibrium spin polarization effect and study its dependence on various ingredients in the modeling. To quantify this effect, we will use y component $P_{y}$ of the spin polarization vector and y component $S_{y}(t)$ of the average spin vector, which have been defined in Eq.(\ref{eq:sp}) and Eq.(\ref{eq:as}). In our calculations, we found that the x and z components of the spin polarization and the average spin vectors tend to be zero and can be ignored, consistent with the magnetic field source. The spin polarization $P_y$ is shown in Fig.~\ref{fig_SP} as a function of time t. The absolute of $P_y$ monotonically decays with time for all of the quarks, which is consistent with the decay of the magnetic field. However, interactions between quarks will affect the decay behavior of spin polarization. The weaker relaxation time, the stronger interaction. The right picture in Fig.~\ref{fig_SP} shows the comparison of spin polarization about u quark for different relaxation times, and reflects that the interaction will delay the decay of spin polarization at early stage, the stronger of interaction between quarks, the stronger delay. And then the interactions between the quarks accelerate the decay, and the stronger the interaction, the faster the decay. This reason is connected with the average spin vector, which will be described next.

\begin{figure}[!hbt]
	\begin{center}
		\includegraphics[scale=0.32]{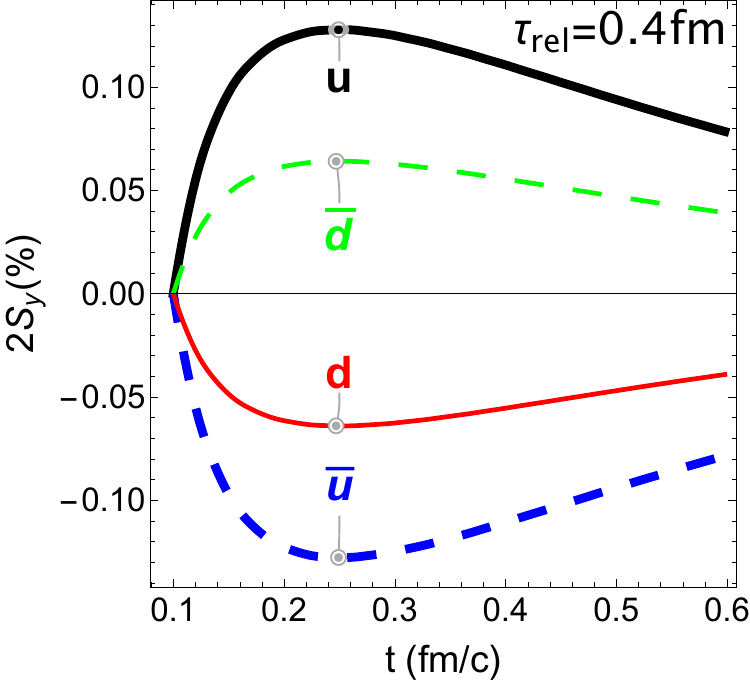}
		\includegraphics[scale=0.32]{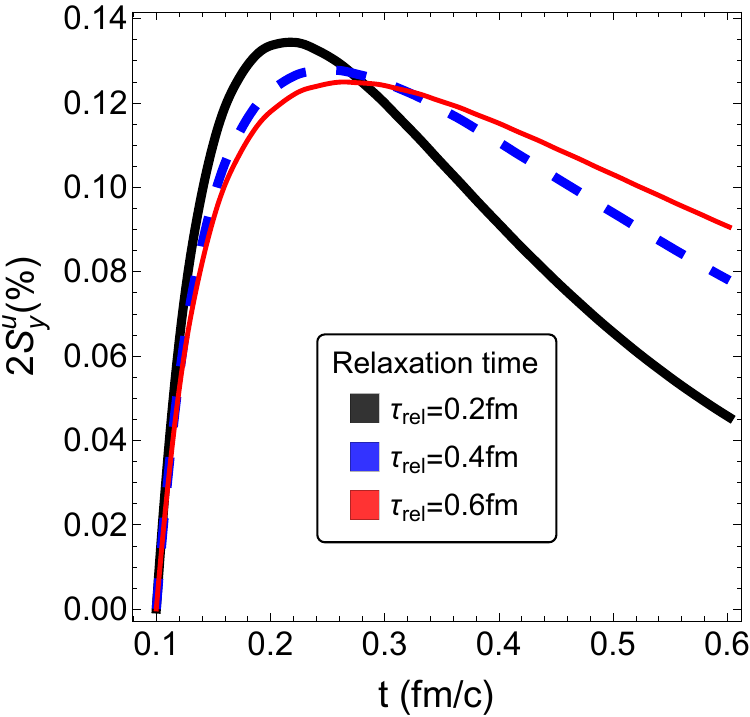}
		\caption{(left)The evolution picture of the average spin of different quarks, computed with the RTA $\tau_{R}=0.4\mathrm{fm}$. (right) Comparing the average spin of u quark for different RTA.}
		\vspace{-0.5cm}
		\label{fig_AS}
	\end{center}
\end{figure}

It will be more explicit about the effect of interaction between quarks on the average spin vector. The average spin $S_y$ as a function of time is shown in Fig.~\ref{fig_AS}. It shows the absolute average spin for all of the quarks monotonically increases to a peak around time $t=0.2\rm fm$, and then monotonically decays with time. As shown in the right part in Fig.~\ref{fig_AS}, the average spin is very sensitive to the interaction between quarks. The stronger interaction between quarks, the faster increase to peak at an early time, and then faster decay from the peak. It is clear that the interaction between quarks on the average spin in the fireball is an accelerating effect, accelerating not only its growth but also its decay. The interaction between quarks will enhance the polarization of spin of quarks from zero average spin stage, and meantime it will destroy the polarization of spin. The average spin vector is the zeroth order part of spin polarization defined in Eq.(\ref{eq:sp}) as mentioned before. Then one can be understood the reason of the behavior of spin polarization in the right-hand picture of Fig.~\ref{fig_SP}.  

We want to notice that the difference of the average spin (or spin polarization) for quark and anti-quark scaled by the quark charge is the same at any time during the evolution, i.e ${\Delta}S^{u}_y(t)/q_u={\Delta}S^{d}_y(t)/q_d$, or ${\Delta}P^{u}_y(t)/q_u={\Delta}P^{d}_y(t)/q_d$, where ${\Delta}S^{u}_y=S^{u}_y-S^{\bar{u}}_y$ and ${\Delta}P^{u}_y=P^{u}_y-P^{\bar{u}}_y$. 

\begin{figure}[!hbt]
	\begin{center}
		\includegraphics[scale=0.32]{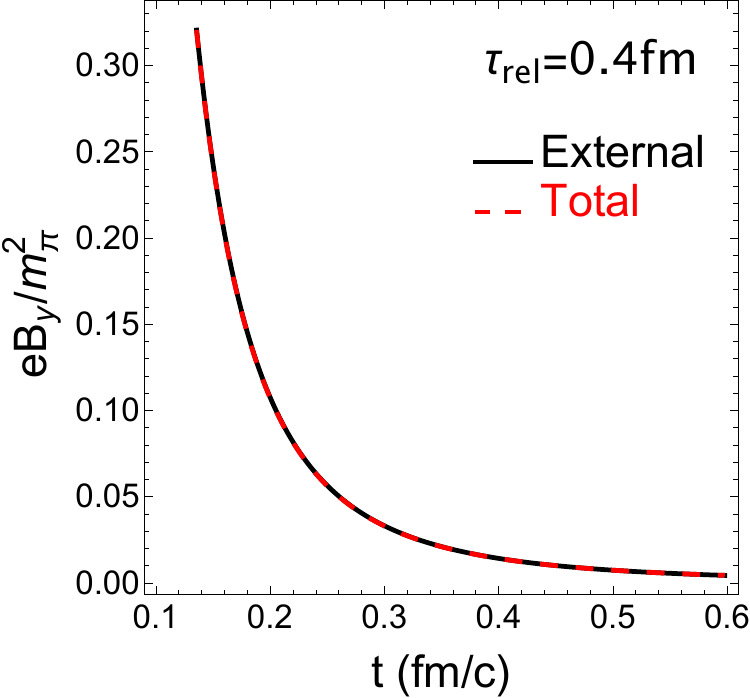}
		\includegraphics[scale=0.32]{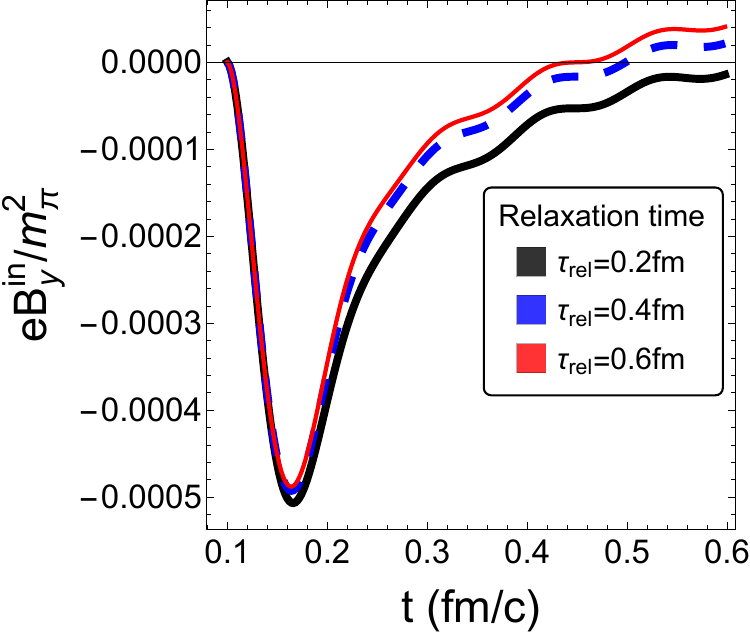}
		\caption{(left)The evolution picture of the external and total magnetic field at the center of the fireball, computed with the RTA $\tau_{R}=0.4\mathrm{fm}$. (right) Comparing the induced (internal) magnetic field for different RTA.}
		\vspace{-0.5cm}
		\label{fig_B}
	\end{center}
\end{figure}

We next examine the influence of the interaction between quarks on the dynamical evolution of the magnetic field at an early stage in heavy-ion collisions. We plot the comparison of the external magnetic field $B^{ext}_y(t)$ and total magnetic field $B^{total}_y(t)=B^{ext}_y(t)+B^{int}_y(t)$ as a function of time computed for RTA $\tau_{R}=0.4\rm fm$ shown in the left part in Fig.~\ref{fig_B}. The external and total magnetic fields are almost identical, a result consistent with previous works\cite{Voronyuk:2011jd,Zakharov:2014dia, Wang:2021oqq}. This means that the induced (internal) magnetic field is very small at the early stage in heavy-ion collisions, and can be negligible compared to the external magnetic field. However, this conclusion does not prevent further studies on the evolution of magnetic fields in the early stages of heavy ion collisions. In this work, we have not yet considered some key factors, namely the net charge effect and the vorticity effect, both of which have a large effect on the induced magnetic field\cite{Guo:2019mgh}.

Although the induced magnetic field is very small, we also provided a detailed depiction of its dynamical evolution and found that the induced magnetic field is not significantly affected by the interactions between quarks, as illustrated in the right panel of Figure.~\ref{fig_B}. Our results indicate an unusual electromagnetic response from the fireball. The induced magnetic field initially decreases with the decay of the external magnetic field in the y-direction, then increases after reaching its nadir. Generally, according to Lenz's law, the induced magnetic field should be positive and rise rapidly to a 'stable' value. This unusual response may arise from the quantum correction induced by the Berry terms (which include the Berry curvature $\mathbf{b}_{\chi}$) in the equation of motion of chiral transport equation Eq.(\ref{eq:eom}). 

\begin{figure}[!hbt]
	\begin{center}
		\includegraphics[scale=0.4]{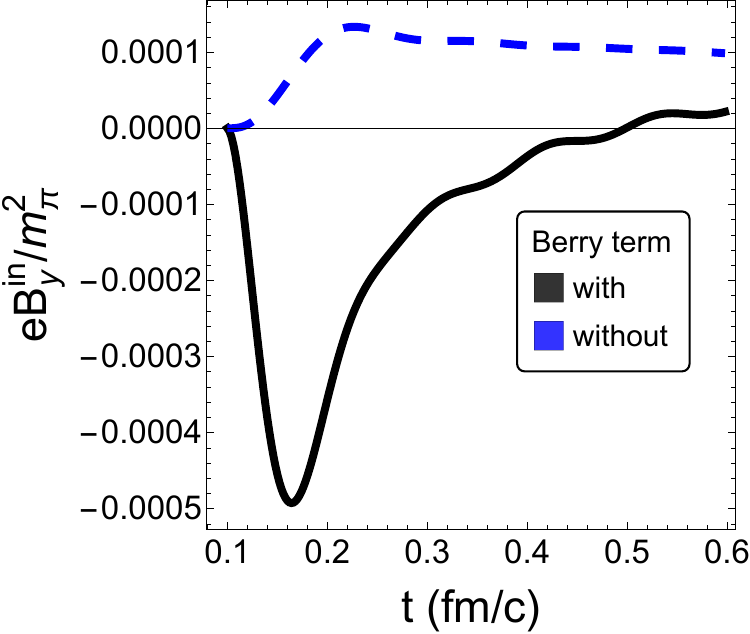}
		\caption{The induced (internal) magnetic field in the case with and without the Berry term for the Relaxation Time Approximation (RTA) $\tau_{R}=0.4\mathrm{fm}$ .}
		\vspace{-0.5cm}
		\label{fig_B_Berry}
	\end{center}
\end{figure}

To clarify this phenomenon, we omit all Berry terms from the equation of motion   Eq.(\ref{eq:eom}), thereby transforming the chiral transport equation into a conventional kinetic equation devoid of quantum corrections. This allows the massless quarks to behave as classical particles. We observe that, in the absence of quantum corrections, the induced magnetic field increases as the decay of the external magnetic field in the y-direction, as depicted in Figure.~\ref{fig_B_Berry}. This finding aligns with our previous assumption. Hence, the unusual response of the internal magnetic field can be attributed to quantum corrections. This represents an incomplete electromagnetic response effect, distinct from the response described by Lenz's law. In this case, the induced magnetic field quickly increases in the opposite direction to that predicted by Lenz's law, followed by a rapid increase in the direction aligned with Lenz's law. A key point to note is that there is no spin polarization in this conventional kinetic equation. This also serves to verify the validity of our code from another perspective.

\begin{figure}[!hbt]
	\begin{center}
		\includegraphics[scale=0.32]{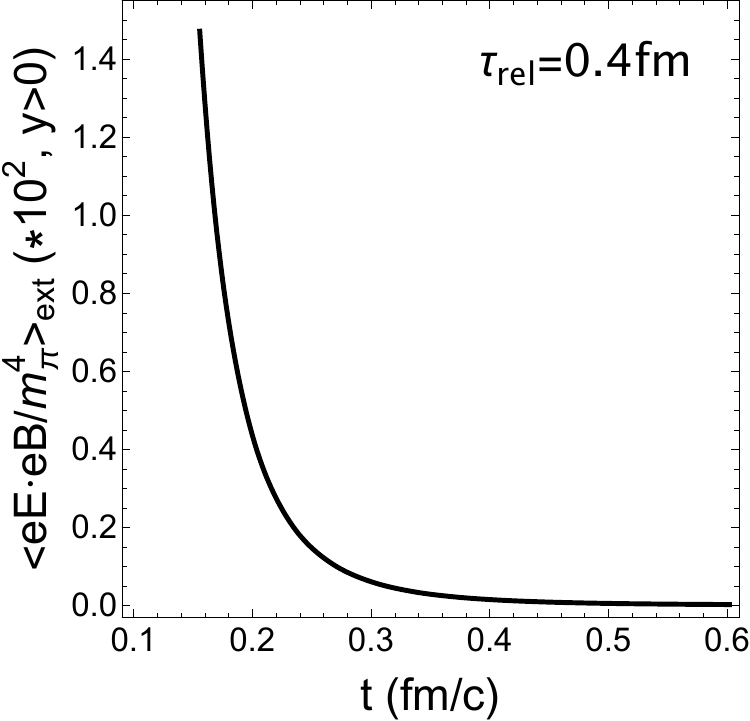}
		\includegraphics[scale=0.32]{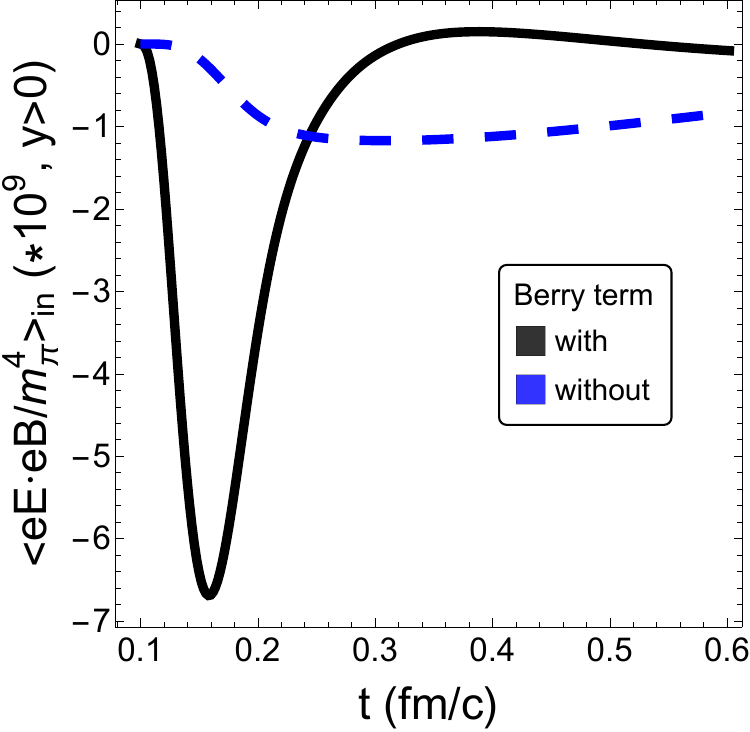}
		\caption{(left) The evolution plot of the average external $\langle e^2\mathbf{E}\cdot \mathbf{B}\rangle$ normalized to the pion mass, as computed using the Relaxation Time Approximation (RTA) with $\tau_{R}=0.4\mathrm{fm}$. (right) A comparison of the induced average $\langle e^2\mathbf{E}\cdot \mathbf{B}\rangle$ normalized to the pion mass, for scenarios with and without the inclusion of the Berry term, under the RTA $\tau_{R}=0.4\mathrm{fm}$ .}
		\vspace{-0.5cm}
		\label{fig_EB}
	\end{center}
\end{figure}

The unusual response is also evident in the inner product of the induced electric and magnetic fields, $\langle e^2\mathbf{E}\cdot \mathbf{B}\rangle$. The average over the simulation is given by $\langle e^2\mathbf{E}\cdot \mathbf{B}\rangle=\int d^3x n(x) e^2\mathbf{E}\cdot \mathbf{B}/\int d^3x n(x)$, where $n(x)$ represents the quarks number density. As depicted in the right panel of Figure.~\ref{fig_EB}, the evolution of the inner product of the induced electric and magnetic fields exhibits distinct differences between the cases with and without Berry terms. This discrepancy is attributed to quantum corrections introduced by the Berry terms, the essence of which lies in the interaction between the spin of quarks and electromagnetic fields. To compare the magnitude of the inner product of the induced and external fields, we also illustrate the inner product of the external electric and magnetic fields as a function of time on the left side of Figure~\ref{fig_EB} for the Relaxation Time Approximation with $\tau_{R}=0.4\mathrm{fm}$.

Lastly, it is important to note that the induced or external electric fields are close to zero at the center of fireball for all the above cases and, therefore, are not shown in this presentation.



\section{Summary}
\label{summary}
In summary, we have performed a phenomenological study of the magnetic-induced spin polarization of quarks and the dynamical magnetic field during the pre-thermal stage in heavy ion collisions. In such a collision, the extremely strong magnetic field may last only for a brief moment and the magnetic-induced spin polarization may occur at so early a stage that the quark-gluon matter is still far from thermal equilibrium. Utilizing the tool of chiral kinetic theory, we have shown the dynamical evolution of magnetic-induced spin polarization of QGP by the spin polarization and average spin of quarks and studied its dependence on various ingredients in the modeling. The effect is found to be sensitive to the relaxation time of the system evolution toward thermal equilibrium. In the present work, it is found that this pre-thermal magnetic-induced spin polarization is considerable. 

In our simulation, we find that the magnetic-induced spin polarization monotonically decreases as the magnetic field decays. Interestingly, the interaction between quarks can delay the decay of spin polarization at an early stage but accelerate it later. However, the magnetic-induced average spin monotonically increases to a peak at first and then monotonically decays with time. The effect of interaction between quarks on the magnetic-induced average spin is an accelerating effect that not only accelerates its growth but also its decay. It enhances the average spin from zero at first and then accelerates its destruction later. 

We have also investigated the dynamical evolution of the induced magnetic field within a pre-thermal Quark-Gluon Plasma (QGP), examining its dependence on various relaxation times and the impact of quantum corrections. The induced magnetic field is generally much weaker than the external magnetic field. It remains insensitive to quark interactions but is significantly influenced by quantum corrections arising from Berry terms. Moreover, the electromagnetic response of the fireball exhibits unusual characteristics, reverting to a more conventional behavior in the absence of Berry's terms. We predict that this unusual response is a universal feature of non-equilibrium quantum systems and may also manifest in massive systems. Therefore, one could explore this unique response in systems composed of massive charged particles with spin under a decaying magnetic field.

The current study has inspired us to delve deeper into the dynamics of spin polarization and magnetic fields in lower-energy heavy-ion collisions. In such collisions, the external magnetic field decays more slowly due to the relatively slower movement of the projectile and target. Consequently, quarks experience a prolonged magnetic-induced spin polarization, which presents a unique opportunity for us to explore this phenomenon in greater detail. Based on the present work, we intend to develop a more comprehensive and realistic framework to study the dynamic evolution of the magnetic field and spin polarization in low-energy heavy-ion collisions in the future. This future work will be conducted by replacing the kinetic theory with chiral kinetic theory and incorporating Maxwell equations into the ZPC.f program of A Multi-Phase Transport(AMPT)\cite{Zhang:1999bd,Lin:2001zk,Lin:2002gc}, which is employed to simulate the Parton cascade.  This will allow us to delve deeper into this fascinating phenomenon and gain a better understanding of its underlying mechanisms.


\vspace{0.2in}

 {\bf Acknowledgments.} 
The research of A.H. especially thanks to the support of the National Natural Science Foundation of China (NSFC) Grant No.12205309. The work is also supported by the the National Natural Science Foundation of China (NSFC) Grant No. 12235016 and No. 12221005, the Strategic Priority Research Program of Chinese Academy of Sciences Grant No. XDB34030000, the Fundamental Research Funds for the Central Universities.

\bibliography{reference}

\begin{thebibliography}{10}

\bibitem{Kharzeev:2004ey}
Dmitri Kharzeev.
\newblock {Parity violation in hot QCD: Why it can happen, and how to look for
  it}.
\newblock {\em Phys. Lett. B}, 633:260--264, 2006.

\bibitem{Kharzeev:2007tn}
D.~Kharzeev and A.~Zhitnitsky.
\newblock {Charge separation induced by P-odd bubbles in QCD matter}.
\newblock {\em Nucl. Phys. A}, 797:67--79, 2007.

\bibitem{Kharzeev:2007jp}
Dmitri~E. Kharzeev, Larry~D. McLerran, and Harmen~J. Warringa.
\newblock {The Effects of topological charge change in heavy ion collisions:
  'Event by event P and CP violation'}.
\newblock {\em Nucl. Phys. A}, 803:227--253, 2008.

\bibitem{Fukushima:2008xe}
Kenji Fukushima, Dmitri~E. Kharzeev, and Harmen~J. Warringa.
\newblock {The Chiral Magnetic Effect}.
\newblock {\em Phys. Rev. D}, 78:074033, 2008.

\bibitem{Kharzeev:2010gd}
Dmitri~E. Kharzeev and Ho-Ung Yee.
\newblock {Chiral Magnetic Wave}.
\newblock {\em Phys. Rev. D}, 83:085007, 2011.

\bibitem{Burnier:2011bf}
Yannis Burnier, Dmitri~E. Kharzeev, Jinfeng Liao, and Ho-Ung Yee.
\newblock {Chiral magnetic wave at finite baryon density and the electric
  quadrupole moment of quark-gluon plasma in heavy ion collisions}.
\newblock {\em Phys. Rev. Lett.}, 107:052303, 2011.

\bibitem{Son:2009tf}
Dam~T. Son and Piotr Surowka.
\newblock {Hydrodynamics with Triangle Anomalies}.
\newblock {\em Phys. Rev. Lett.}, 103:191601, 2009.

\bibitem{Kharzeev:2010gr}
Dmitri~E. Kharzeev and Dam~T. Son.
\newblock {Testing the chiral magnetic and chiral vortical effects in heavy ion
  collisions}.
\newblock {\em Phys. Rev. Lett.}, 106:062301, 2011.

\bibitem{Sadofyev:2010is}
A.~V. Sadofyev, V.~I. Shevchenko, and V.~I. Zakharov.
\newblock {Notes on chiral hydrodynamics within effective theory approach}.
\newblock {\em Phys. Rev. D}, 83:105025, 2011.

\bibitem{Landsteiner:2011iq}
Karl Landsteiner, Eugenio Megias, Luis Melgar, and Francisco Pena-Benitez.
\newblock {Holographic Gravitational Anomaly and Chiral Vortical Effect}.
\newblock {\em JHEP}, 09:121, 2011.

\bibitem{Jiang:2015cva}
Yin Jiang, Xu-Guang Huang, and Jinfeng Liao.
\newblock {Chiral vortical wave and induced flavor charge transport in a
  rotating quark-gluon plasma}.
\newblock {\em Phys. Rev. D}, 92(7):071501, 2015.

\bibitem{STAR:2017ckg}
L.~Adamczyk et~al.
\newblock {Global $\Lambda$ hyperon polarization in nuclear collisions:
  evidence for the most vortical fluid}.
\newblock {\em Nature}, 548:62--65, 2017.

\bibitem{STAR:2018gyt}
Jaroslav Adam et~al.
\newblock {Global polarization of $\Lambda$ hyperons in Au+Au collisions at
  $\sqrt{s_{_{NN}}}$ = 200 GeV}.
\newblock {\em Phys. Rev. C}, 98:014910, 2018.

\bibitem{STAR:2019erd}
Jaroslav Adam et~al.
\newblock {Polarization of $\Lambda$ ($\bar{\Lambda}$) hyperons along the beam
  direction in Au+Au collisions at $\sqrt{s_{_{NN}}}$ = 200 GeV}.
\newblock {\em Phys. Rev. Lett.}, 123(13):132301, 2019.

\bibitem{Niida:2018hfw}
Takafumi Niida.
\newblock {Global and local polarization of $\Lambda$ hyperons in Au+Au
  collisions at 200 GeV from STAR}.
\newblock {\em Nucl. Phys. A}, 982:511--514, 2019.

\bibitem{ALICE:2019onw}
Shreyasi Acharya et~al.
\newblock {Global polarization of $\Lambda \bar \Lambda$ hyperons in Pb-Pb
  collisions at $\sqrt {s_{NN}}$ = 2.76 and 5.02 TeV}.
\newblock {\em Phys. Rev. C}, 101(4):044611, 2020.
\newblock [Erratum: Phys.Rev.C 105, 029902 (2022)].

\bibitem{Becattini:2013fla}
F.~Becattini, V.~Chandra, L.~Del~Zanna, and E.~Grossi.
\newblock {Relativistic distribution function for particles with spin at local
  thermodynamical equilibrium}.
\newblock {\em Annals Phys.}, 338:32--49, 2013.

\bibitem{Fang:2016vpj}
Ren-hong Fang, Long-gang Pang, Qun Wang, and Xin-nian Wang.
\newblock {Polarization of massive fermions in a vortical fluid}.
\newblock {\em Phys. Rev. C}, 94(2):024904, 2016.

\bibitem{Pang:2016igs}
Long-Gang Pang, Hannah Petersen, Qun Wang, and Xin-Nian Wang.
\newblock {Vortical Fluid and $\Lambda$ Spin Correlations in High-Energy
  Heavy-Ion Collisions}.
\newblock {\em Phys. Rev. Lett.}, 117(19):192301, 2016.

\bibitem{Liu:2019krs}
Shuai Y.~F. Liu, Yifeng Sun, and Che~Ming Ko.
\newblock {Spin Polarizations in a Covariant Angular-Momentum-Conserved Chiral
  Transport Model}.
\newblock {\em Phys. Rev. Lett.}, 125(6):062301, 2020.

\bibitem{Wu:2019eyi}
Hong-Zhong Wu, Long-Gang Pang, Xu-Guang Huang, and Qun Wang.
\newblock {Local spin polarization in high energy heavy ion collisions}.
\newblock {\em Phys. Rev. Research.}, 1:033058, 2019.

\bibitem{Florkowski:2019voj}
Wojciech Florkowski, Avdhesh Kumar, Radoslaw Ryblewski, and Aleksas
  Mazeliauskas.
\newblock {Longitudinal spin polarization in a thermal model}.
\newblock {\em Phys. Rev. C}, 100(5):054907, 2019.

\bibitem{Fu:2021pok}
Baochi Fu, Shuai Y.~F. Liu, Longgang Pang, Huichao Song, and Yi~Yin.
\newblock {Shear-Induced Spin Polarization in Heavy-Ion Collisions}.
\newblock {\em Phys. Rev. Lett.}, 127(14):142301, 2021.

\bibitem{Fu:2020oxj}
Baochi Fu, Kai Xu, Xu-Guang Huang, and Huichao Song.
\newblock {Hydrodynamic study of hyperon spin polarization in relativistic
  heavy ion collisions}.
\newblock {\em Phys. Rev. C}, 103(2):024903, 2021.

\bibitem{Muller:2018ibh}
Berndt M\"uller and Andreas Sch\"afer.
\newblock {Chiral magnetic effect and an experimental bound on the late time
  magnetic field strength}.
\newblock {\em Phys. Rev. D}, 98(7):071902, 2018.

\bibitem{Guo:2019mgh}
Xingyu Guo, Jinfeng Liao, and Enke Wang.
\newblock {Spin Hydrodynamic Generation in the Charged Subatomic Swirl}.
\newblock {\em Sci. Rep.}, 10(1):2196, 2020.

\bibitem{Guo:2019joy}
Yu~Guo, Shuzhe Shi, Shengqin Feng, and Jinfeng Liao.
\newblock {Magnetic Field Induced Polarization Difference between Hyperons and
  Anti-hyperons}.
\newblock {\em Phys. Lett. B}, 798:134929, 2019.

\bibitem{Han:2019fce}
Zhang-Zhu Han and Jun Xu.
\newblock {Charge asymmetry dependence of the elliptic flow splitting in
  relativistic heavy-ion collisions}.
\newblock {\em Phys. Rev. C}, 99(4):044915, 2019.

\bibitem{Becattini:2020ngo}
Francesco Becattini and Michael~A. Lisa.
\newblock {Polarization and Vorticity in the Quark\textendash{}Gluon Plasma}.
\newblock {\em Ann. Rev. Nucl. Part. Sci.}, 70:395--423, 2020.

\bibitem{Xu:2022hql}
Kun Xu, Fan Lin, Anping Huang, and Mei Huang.
\newblock {\ensuremath{\Lambda}/\ensuremath{\Lambda}\textasciimacron{}
  polarization and splitting induced by rotation and magnetic field}.
\newblock {\em Phys. Rev. D}, 106(7):L071502, 2022.

\bibitem{Stephanov:2012ki}
M.~A. Stephanov and Y.~Yin.
\newblock {Chiral Kinetic Theory}.
\newblock {\em Phys. Rev. Lett.}, 109:162001, 2012.

\bibitem{Son:2012wh}
Dam~Thanh Son and Naoki Yamamoto.
\newblock {Berry Curvature, Triangle Anomalies, and the Chiral Magnetic Effect
  in Fermi Liquids}.
\newblock {\em Phys. Rev. Lett.}, 109:181602, 2012.

\bibitem{Son:2012zy}
Dam~Thanh Son and Naoki Yamamoto.
\newblock {Kinetic theory with Berry curvature from quantum field theories}.
\newblock {\em Phys. Rev. D}, 87(8):085016, 2013.

\bibitem{Chen:2012ca}
Jiunn-Wei Chen, Shi Pu, Qun Wang, and Xin-Nian Wang.
\newblock {Berry Curvature and Four-Dimensional Monopoles in the Relativistic
  Chiral Kinetic Equation}.
\newblock {\em Phys. Rev. Lett.}, 110(26):262301, 2013.

\bibitem{Kharzeev:2016sut}
Dmitri~E. Kharzeev, Mikhail~A. Stephanov, and Ho-Ung Yee.
\newblock {Anatomy of chiral magnetic effect in and out of equilibrium}.
\newblock {\em Phys. Rev. D}, 95(5):051901, 2017.

\bibitem{Hidaka:2016yjf}
Yoshimasa Hidaka, Shi Pu, and Di-Lun Yang.
\newblock {Relativistic Chiral Kinetic Theory from Quantum Field Theories}.
\newblock {\em Phys. Rev. D}, 95(9):091901, 2017.

\bibitem{Mueller:2017arw}
Niklas Mueller and Raju Venugopalan.
\newblock {Worldline construction of a covariant chiral kinetic theory}.
\newblock {\em Phys. Rev. D}, 96(1):016023, 2017.

\bibitem{Gorbar:2017cwv}
E.~V. Gorbar, V.~A. Miransky, I.~A. Shovkovy, and P.~O. Sukhachov.
\newblock {Second-order chiral kinetic theory: Chiral magnetic and
  pseudomagnetic waves}.
\newblock {\em Phys. Rev. B}, 95(20):205141, 2017.

\bibitem{Huang:2018wdl}
Anping Huang, Shuzhe Shi, Yin Jiang, Jinfeng Liao, and Pengfei Zhuang.
\newblock {Complete and Consistent Chiral Transport from Wigner Function
  Formalism}.
\newblock {\em Phys. Rev. D}, 98(3):036010, 2018.

\bibitem{Chen:2015gta}
Jing-Yuan Chen, Dam~T. Son, and Mikhail~A. Stephanov.
\newblock {Collisions in Chiral Kinetic Theory}.
\newblock {\em Phys. Rev. Lett.}, 115(2):021601, 2015.

\bibitem{Huang:2017tsq}
Anping Huang, Yin Jiang, Shuzhe Shi, Jinfeng Liao, and Pengfei Zhuang.
\newblock {Out-of-equilibrium chiral magnetic effect from chiral kinetic
  theory}.
\newblock {\em Phys. Lett. B}, 777:177--183, 2018.

\bibitem{Gorbar:2016qfh}
E.~V. Gorbar, I.~A. Shovkovy, S.~Vilchinskii, I.~Rudenok, A.~Boyarsky, and
  O.~Ruchayskiy.
\newblock {Anomalous Maxwell equations for inhomogeneous chiral plasma}.
\newblock {\em Phys. Rev. D}, 93(10):105028, 2016.

\bibitem{Yang:2020hri}
Di-Lun Yang, Koichi Hattori, and Yoshimasa Hidaka.
\newblock {Effective quantum kinetic theory for spin transport of fermions with
  collsional effects}.
\newblock {\em JHEP}, 07:070, 2020.

\bibitem{Yamamoto:2023okm}
Naoki Yamamoto and Di-Lun Yang.
\newblock {Chiral kinetic theory with self-energy corrections and neutrino spin
  Hall effect}.
\newblock {\em Phys. Rev. D}, 109(5):056010, 2024.

\bibitem{mclerran2014comments}
L~McLerran and V~Skokov.
\newblock Comments about the electromagnetic field in heavy-ion collisions.
\newblock {\em Nuclear Physics A}, 929:184--190, 2014.

\bibitem{Huang:2022qdn}
Anping Huang, Duan She, Shuzhe Shi, Mei Huang, and Jinfeng Liao.
\newblock {Dynamical magnetic fields in heavy-ion collisions}.
\newblock {\em Phys. Rev. C}, 107(3):034901, 2023.

\bibitem{Han:2017hdi}
Zhang-Zhu Han and Jun Xu.
\newblock {Investigating different $\Lambda$ and $\bar\Lambda$ polarizations in
  relativistic heavy-ion collisions}.
\newblock {\em Phys. Lett. B}, 786:255--259, 2018.

\bibitem{Becattini:2021suc}
F.~Becattini, M.~Buzzegoli, and A.~Palermo.
\newblock {Spin-thermal shear coupling in a relativistic fluid}.
\newblock {\em Phys. Lett. B}, 820:136519, 2021.

\bibitem{Becattini:2020sww}
Francesco Becattini.
\newblock {Polarization in Relativistic Fluids: A Quantum Field Theoretical
  Derivation}.
\newblock {\em Lect. Notes Phys.}, 987:15--52, 2021.

\bibitem{Liu:2021nyg}
Yu-Chen Liu and Xu-Guang Huang.
\newblock {Spin polarization formula for Dirac fermions at local equilibrium}.
\newblock {\em Sci. China Phys. Mech. Astron.}, 65(7):272011, 2022.

\bibitem{Fang:2024vds}
Shuo Fang and Shi Pu.
\newblock {Collisional corrections to spin polarization from quantum kinetic
  theory using Chapman-Enskog expansion}.
\newblock 8 2024.

\bibitem{aristov2001direct}
Vasili{\u\i} Aristov.
\newblock {\em Direct methods for solving the Boltzmann equation and study of
  nonequilibrium flows}.

\bibitem{yee1966numerical}
Kane Yee.
\newblock Numerical solution of initial boundary value problems involving
  maxwell's equations in isotropic media.
\newblock {\em IEEE Transactions on antennas and propagation}, 14(3):302--307,
  1966.

\bibitem{bird1994molecular}
Graeme~A Bird.
\newblock Molecular gas dynamics and the direct simulation of gas flows.
\newblock {\em Molecular gas dynamics and the direct simulation of gas flows},
  1994.

\bibitem{grigoryev2012numerical}
Yu~N Grigoryev and Vitali{\u\i} Vshivkov.
\newblock Numerical" particle-in-cell" methods.

\bibitem{McLerran:2013hla}
L.~McLerran and V.~Skokov.
\newblock {Comments About the Electromagnetic Field in Heavy-Ion Collisions}.
\newblock {\em Nucl. Phys.}, A929:184--190, 2014.

\bibitem{Zakharov:2014dia}
B.~G. Zakharov.
\newblock {Electromagnetic response of quark\textendash{}gluon plasma in
  heavy-ion collisions}.
\newblock {\em Phys. Lett. B}, 737:262--266, 2014.

\bibitem{alver2008phobos}
B~Alver, M~Baker, C~Loizides, and P~Steinberg.
\newblock The phobos glauber monte carlo.
\newblock {\em arXiv preprint arXiv:0805.4411}, 2008.

\bibitem{Yee:1966}
{Kane Yee}.
\newblock Numerical solution of initial boundary value problems involving
  maxwell's equations in isotropic media.
\newblock {\em IEEE Transactions on Antennas and Propagation}, 14(3):302--307,
  1966.

\bibitem{Kowalski:2007rw}
H.~Kowalski, T.~Lappi, and R.~Venugopalan.
\newblock {Nuclear enhancement of universal dynamics of high parton densities}.
\newblock {\em Phys. Rev. Lett.}, 100:022303, 2008.

\bibitem{blaizot2012bose}
Jean-Paul Blaizot, Fran{\c{c}}ois Gelis, Jinfeng Liao, Larry McLerran, and Raju
  Venugopalan.
\newblock Bose--einstein condensation and thermalization of the quark--gluon
  plasma.
\newblock {\em Nuclear Physics A}, 873:68--80, 2012.

\bibitem{blaizot2013gluon}
Jean-Paul Blaizot, Jinfeng Liao, and Larry McLerran.
\newblock Gluon transport equation in the small angle approximation and the
  onset of bose--einstein condensation.
\newblock {\em Nuclear Physics A}, 920:58--77, 2013.

\bibitem{blaizot2014quark}
Jean-Paul Blaizot, Bin Wu, and Li~Yan.
\newblock Quark production, bose--einstein condensates and thermalization of
  the quark--gluon plasma.
\newblock {\em Nuclear Physics A}, 930:139--162, 2014.

\bibitem{Voronyuk:2011jd}
V.~Voronyuk, V.~D. Toneev, W.~Cassing, E.~L. Bratkovskaya, V.~P. Konchakovski,
  and S.~A. Voloshin.
\newblock {(Electro-)Magnetic field evolution in relativistic heavy-ion
  collisions}.
\newblock {\em Phys. Rev. C}, 83:054911, 2011.

\bibitem{Wang:2021oqq}
Zeyan Wang, Jiaxing Zhao, Carsten Greiner, Zhe Xu, and Pengfei Zhuang.
\newblock {Incomplete electromagnetic response of hot QCD matter}.
\newblock {\em Phys. Rev. C}, 105(4):L041901, 2022.

\bibitem{Zhang:1999bd}
Bin Zhang, C.~M. Ko, Bao-An Li, and Zi-wei Lin.
\newblock {A multiphase transport model for nuclear collisions at RHIC}.
\newblock {\em Phys. Rev. C}, 61:067901, 2000.

\bibitem{Lin:2001zk}
Zi-wei Lin and C.~M. Ko.
\newblock {Partonic effects on the elliptic flow at RHIC}.
\newblock {\em Phys. Rev. C}, 65:034904, 2002.

\bibitem{Lin:2002gc}
Zi-wei Lin, C.~M. Ko, and Subrata Pal.
\newblock {Partonic effects on pion interferometry at RHIC}.
\newblock {\em Phys. Rev. Lett.}, 89:152301, 2002.

\end{thebibliography}
\bibliographystyle{unsrt}

\end{document}